\documentclass[a4paper,12pt]{article}
\usepackage{mathrsfs,graphicx,rotating,amsmath,amsfonts,mathtools,booktabs,amssymb}
\usepackage[symbol]{footmisc}
\usepackage{hyperref}\usepackage{slashed}
\usepackage[nosort]{cite}
\usepackage[table,xcdraw]{xcolor}
\hypersetup{colorlinks,bookmarksopen,bookmarksnumbered,
linkcolor=blus,pdfstartview=FitH,urlcolor=rossos,citecolor=verde}
\allowdisplaybreaks

\usepackage{multicol}
\usepackage{color}
\definecolor{rosso}{cmyk}{0,1,1,0.4}
\definecolor{rossos}{cmyk}{0,1,1,0.55}
\definecolor{rossoc}{cmyk}{0,1,1,0.2}
\definecolor{blu}{cmyk}{1,1,0,0.3}
\definecolor{blus}{cmyk}{1,1,0,0.6}
\definecolor{bluc}{cmyk}{1,1,0,0.1}
\definecolor{verde}{cmyk}{0.92,0,0.59,0.25}
\definecolor{verdec}{cmyk}{0.92,0,0.59,0.15}
\definecolor{verdes}{cmyk}{0.92,0,0.59,0.4}

\newcommand{\s}{\,{\rm s}}
\newcommand{\eV}{\,{\rm eV}}

\newcommand{\MeV}{\,{\rm MeV}}
\newcommand{\GeV}{\,{\rm GeV}}
\newcommand{\TeV}{\,{\rm TeV}}
\newcommand{\cm}{\,{\rm cm}}

\newcommand{\K}{\,{\rm K}}
\newcommand{\mK}{\,{\rm mK}}

\def\circa#1{\,\raise.3ex\hbox{$#1$\kern-.75em\lower1ex\hbox{$\sim$}}\,}

\newcommand{\be}{\begin{equation}}
\newcommand{\ee}{\end{equation}}

\font\tenrsfs=rsfs10 at 12pt
\font\sevenrsfs=rsfs7
\font\fiversfs=rsfs5
\newfam\rsfsfam
\textfont\rsfsfam=\tenrsfs
\scriptfont\rsfsfam=\sevenrsfs
\scriptscriptfont\rsfsfam=\fiversfs

\newcommand{\Tg}{T_{\rm gas}}
\newcommand{\Pp}{\mathcal{P}_2}
\newcommand{\dep}{\left(\frac{dE}{dVdt}\right)_{\rm deposited}}

\newcommand{\ie}{\emph{i.e.\,}}

\def\circa#1{\,\raise.3ex\hbox{$#1$\kern-.75em\lower1ex\hbox{$\sim$}}\,}
\makeatletter

\def\hhref#1{\href{http://arxiv.org/abs/#1}{arXiv:#1}} 
 
\def\art{\@ifnextchar[{\eart}{\oart}}
\def\eart[#1]#2#3#4#5#6{{\rm #2}, {\em #3 \bf #4} {\rm (#6) #5} ({\em #1})}

\def\article{\@ifnextchar[{\earticle}{\oarticle}}
\def\oarticle#1#2#3#4#5#6{{\rm #1}, {\em ``#6''}, {\rm #2 #3 (#5) #4}}
\def\earticle[#1]#2#3#4#5#6#7{{\rm #2}, {\em ``#7''}, {\rm #3 #4 (#6) #5}  [\hhref{#1}]}
\def\hepart[#1]#2{{\rm #2, \em#1}}
\def\heparticle[#1]#2#3{#2, {\em ``#3''} [\hhref{#1}]}

\begin{document}

\vspace{2cm}

\renewcommand{\thefootnote}{\fnsymbol{footnote}}

\begin{center}
{\Large\LARGE \bf \color{verdes}
Bounds on Dark Matter decay \\
from 21 cm line}\\[1cm]
{
{\bf Andrea Mitridate}\footnote[2]{andrea.mitridate@sns.it} and
{\bf Alessandro Podo} \footnote[8]{alessandro.podo@sns.it} 
\\[1mm]
{\it Scuola Normale Superiore and INFN, Pisa, Italy}
}
\vspace{1cm}

{\large\bf\color{blus} Abstract}
\begin{quote}
\large
The observation of the cosmic 21-cm spectrum can serve as a probe for Dark Matter properties. We point out that the knowledge of the signal amplitude at a given redshift allows one to put conservative bounds on the DM decay rate which are independent of astrophysical parameters.
These limits are valid for the vast majority of DM models, those without extra IGM cooling or additional background radiation. 
Using the experimental results reported by the EDGES collaboration, we derive bounds that are stronger than the ones derived from other CMB observations and competitive with the ones from indirect detection.
\end{quote}

\thispagestyle{empty}
\bigskip

\end{center}

\renewcommand{\thefootnote}{\arabic{footnote}}
\setcounter{footnote}{0}

\section{Introduction}
The 21-cm line is associated with the transition between the spin singlet and triplet hyperfine states of neutral hydrogen. It provides a powerful probe of the high redshift Universe when, after recombination ($z\simeq1000$) and prior to reionization ($z\lesssim 10$), the baryon content of the Universe is mostly in the form of neutral hydrogen and helium. The relative abundance of the two hyperfine levels can be parametrized in terms of the spin temperature, $T_{\rm{S}}$, as: $n_1/n_0\equiv g_1/g_0\exp(-\Delta E/T_{\rm{S}})$ where $g_{1,0}$ are the multiplicities of the two states and $\Delta E=0.068 \K=2\pi/21\cm$ is the hyperfine splitting. The signal can be seen as an absorption or emission feature in contrast to the low-energy tail of the CMB spectrum, according to whether the spin temperature is larger or smaller than the CMB temperature at the relevant cosmic time.

Accounting for cosmological redshift, the 21 cm signal from the cosmic intergalactic medium (IGM) can be observed in the frequency band $\nu = 1420/(1+z) \,\rm{MHz}$. The global sky averaged signal, observed as a function of the frequency, provides a map of the IGM average spin temperature as a function of time, tracing the cosmic history of neutral hydrogen. Working in the Rayleigh-Jeans limit, where the intensity of radiation is proportional to the temperature of the corresponding source, the amplitude of the signal can be parametrized by the brightness temperature offset \cite{1511.01131,Madau:1996cs}:
\be
\delta T_{b}(z) = 27\, x_{\rm{HI}} \left( 1-\frac{T_{\rm{CMB}}(z)}{T_{\rm{S}}(z)}\right) \left( \frac{\Omega_{\rm{b}}h^2}{0.023}\right) \left( \frac{\Omega_{\rm{m}}h^2}{0.15}\right)^{-1/2} \sqrt{\frac{1+z}{10}} \, \rm{mK}\,,
\ee
where $x_{\rm{HI}}$ is the fraction of neutral hydrogen, very close to 1 before reionization, $T_{\rm{CMB}}(z)=2.73\, (1+z) \, \K$ is the CMB temperature at redshift $z$ and $T_{\rm{S}}(z)$ is the spin temperature previously defined. 

The spin temperature is set by three competing processes: scattering with CMB photons, collisional coupling of the IGM and interactions with Lyman-$\alpha$ photons from the first stars \cite{1511.01131}. The first one tends to couple $T_{\rm{S}}$ and $T_{\rm{CMB}}$, while the other two, if efficient, couple $T_{\rm{S}}$ with the kinetic temperature of the gas, $T_{\rm{gas}}$. The gas temperature traces the CMB temperature until Compton scattering becomes inefficient ($z\sim 200$) and the gas starts to cool adiabatically (\emph{i.e.} $T_{gas}\propto(1+z)^2$).
The shape of the global 21-cm cosmic signal is determined by the interplay of these three different processes and by the evolution of the gas temperature. In particular, the presence of an absorption signal is related to the time at which the first stars form and start to recouple the spin temperature to the kinetic temperature of the gas. The IGM can be heated by various astrophysical mechanism among which shock heating and heating from UV and X-rays photons \cite{1511.01131}. Dark Matter (DM) annihilations or decays can be an additional source of gas heating which can suppress or erase the absorption signal expected in the 21 cm spectrum. These mechanisms are sensitive to a number of astrophysical parameters and the observation of the global 21 cm cosmic signal is not expected to be sufficient to disentangle the effect of DM from the uncertainty on these parameters \cite{1209.2120,1408.1109,1603.06795}. Only a statistical study of the spatial fluctuations of the 21 cm cosmic signal will be able to break the degeneracy.

However, we point out that the observation of an absorption line in the global 21 cm spectrum, at a given redshift, allows one to put conservative constraints on the decay of DM particles which are independent of the astrophysical parameters. Whereas the signal shape and position depend on those parameters, the amplitude of an absorption signal at a given redshift is constrained by the condition $T_{\rm{S}}(z) \ge T_{\rm{gas}}(z)$. Since the astrophysical sources can only heat the IGM, a lower bound on $T_{\rm{gas}}(z)$ can be obtained by assuming that its evolution is determined by Compton scattering and adiabatic cooling alone. This results in a conservative  bound on the amplitude of the absorption signal as a function of redshift. The presence of DM decays or annihilations can inject energy in the IGM, heating the gas and resulting in a reduced amplitude of the signal; therefore, the observation of a signal of definite amplitude at a given redshift implies an upper bound on the amount of energy injected in the IGM by DM interactions. The idea to put conservative bounds on energy injection due to annihilation of DM particles has been proposed in \cite{1803.03629}; however, in that case the bounds are sensitive to astrophysical uncertainties (see section~\ref{sec:comparison} for more details). 

\subsection{Discussion of the assumptions}
Recently, the EDGES collaboration reported a detection of an absorption signal in the 21 cm spectrum at redshift $z\approx 17$ with amplitude 
\be
\delta T_{\rm EDGES} = -500^{+200}_{-500}\; \rm{mK}\,,
\ee
with uncertainties at $99 \%$ confidence \cite{Bowman:2018yin}. 

The discrepancy between the value of the amplitude predicted by standard cosmology ($\delta T_b\gtrsim-200\,{\rm mK}$) and the one observed by EDGES has been at the center of many speculations in the recent literature. Many proposals have been put forward to give a physical explanation of the anomaly: modified cosmological evolution, excess of radiation in the tail of the CMB spectrum or DM-baryon scattering \cite{Bowman:2018yin,1509.00029,1803.06698,1802.10094,1803.07048,1803.03245}. 

However, even if the EDGES observation provides evidence for an absorption signal at more than $5\sigma$, the statistical significance of the measurement is not enough to rule out conventional cosmology at the $5\sigma$ level.  We believe that this result provides a strong evidence for the existence of a signal but it is still not enough to unequivocally establish the need of new physics beyond the standard cosmological scenario. 

Therefore, we do not include non-standard sources of cooling of the IGM, excess radiation in the tail of the CMB spectrum or consider the possibility of a modified cosmological evolution. We assume that only standard astrophysical processes are at work and we use the conservative assumption that the signal has an amplitude larger than $-100\; \rm{mK}$ ($-50\; \rm{mK}$), to derive bounds on the DM lifetime.

Given this assumption, the validity of our results is limited to models that do not provide an enhanced absorption signal. The majority of the DM models present in the literature fulfill this assumption\footnote{For a discussion of the features necessary to give an enhanced absorption signal see for instance \cite{1803.02804,1803.03091}.}.

\begin{figure}[t]
\centering
\includegraphics[width=.47\textwidth]{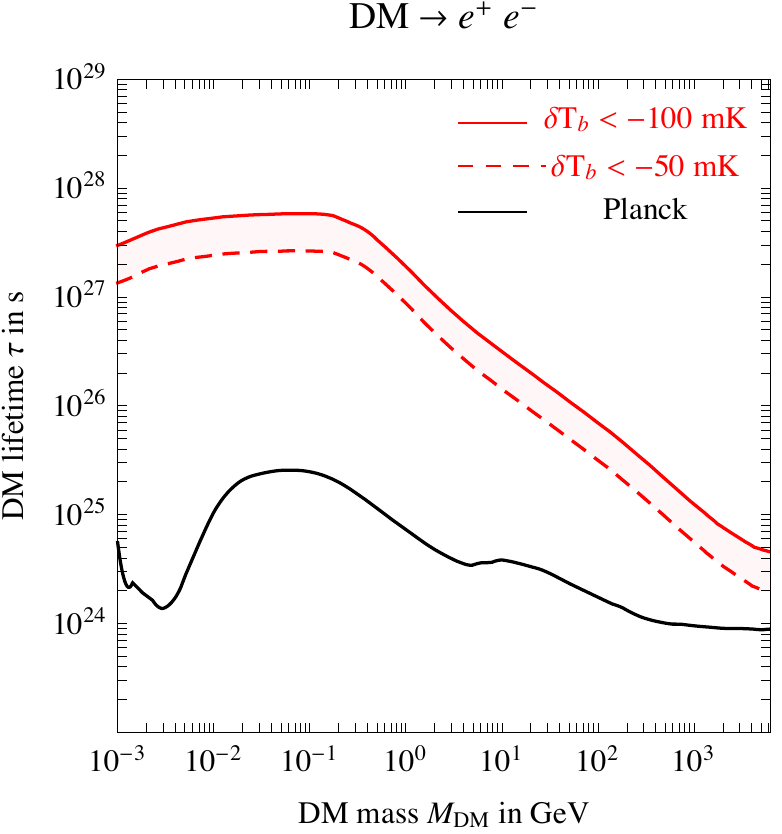}\quad\quad
\includegraphics[width=.47\textwidth]{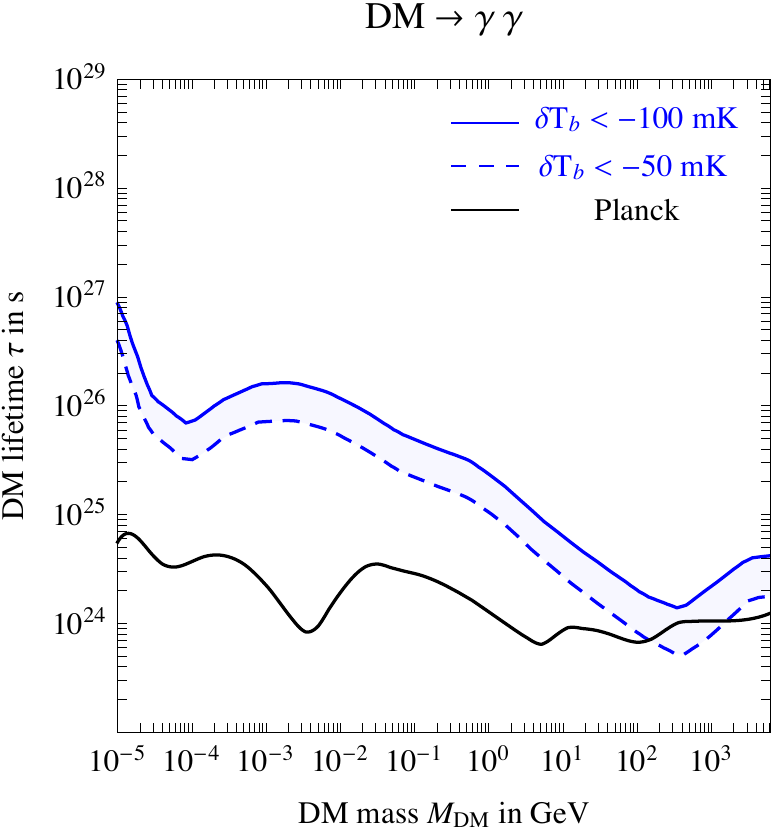}
\caption{\emph{ Bounds on DM lifetime assuming decays into electrons (left panel) or photons (right panel). The solid (dashed) red line is obtained requiring that the amplitude of the absorption signal, including DM decays, satisfies $\delta T_b<-100\,(-50) \mK$ and assuming a single component DM, i.e. taking $f_{\rm DM}=1$. For comparison we report the bounds derived from observations of the CMB power spectrum}\cite{1502.01589,1610.06933}.}
\label{fig:bounds}
\end{figure}

\begin{figure}[t]
\centering
\includegraphics[width=.8\textwidth]{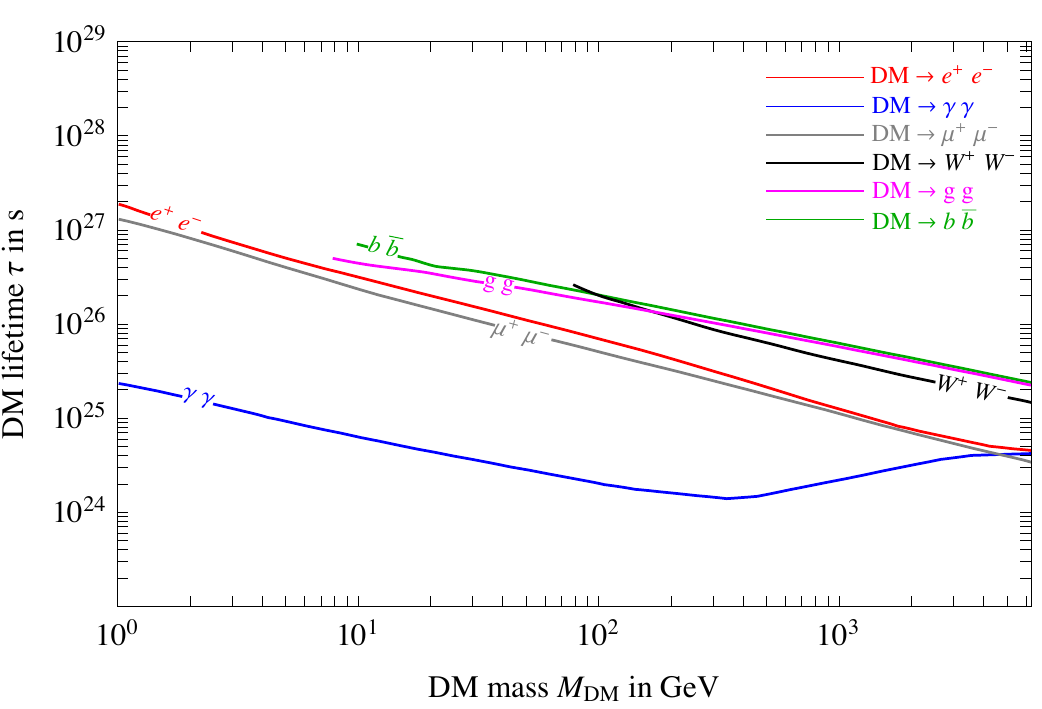}
\caption{\emph{Bounds on DM decaying with a $100\%$ branching fraction in different channels. We require $\delta T_b<-100\mK$ and assume a single component DM, \ie $f_{\rm DM}=1$}.}
\label{fig:all}
\end{figure}
\section{Bounds on Dark Matter decays}
Decays of DM particles inject energy in the Universe at a rate given by:
\be
\label{eq:energy_inj_decay}
\left(\frac{dE}{dVdt}\right)_{\rm injected}=f_{_{\rm DM}} \rho_{_{\rm DM,0}}\, \tau_{_{\rm DM}}^{-1}(1+z)^3 \,,
\ee
where $f_{_{\rm DM}}$ is the fraction of DM that can decay, $\rho_{_{\rm DM,0}}$ is the DM energy density today and $\tau_{_{\rm DM}}$ is the DM lifetime. The injected energy can either be absorbed by the IGM (through ionization and excitation of hydrogen atoms or heating of the gas) or free-stream until today. The reionization history of the Universe is modified only by the fraction of injected energy that is deposited in the IGM. The energy deposition rate per unit volume is given by:
\be 
\left(\frac{dE}{dVdt}\right)_{\rm deposited}\equiv f(z,M_{\rm DM})\left(\frac{dE}{dVdt}\right)_{\rm injected}\,,
\ee
where we have defined the energy deposition efficiency $f(z,M_{\rm DM})$. The $f(z,M_{\rm DM})$ functions also depend on the decay channel, however we omit this dependence for the sake of notation. 

In the presence of this energy injection the evolution of the hydrogen ionization fraction, $x_e\equiv n_e/n_b$, and the gas temperature, $T_{\rm gas}$, is governed by the following equations\cite{Peebles:1968ja,0905.0003,1503.04827}:

\begin{subequations}\label{eq:evol}
\begin{align}\label{eq:evola}
\frac{dx_e}{dz}=&\frac{\Pp}{(1+z)H(z)}\Big[\alpha_{\rm H}\,n_{\rm H}\,x_e^2\, C-\beta_{\rm H}
\,e^{-E_\alpha/T_{\rm CMB}}(1-x_e)\Big]\\ 
\nonumber&-\frac{1}{(1+z)H(z)}\frac{1-x_e}{3n_{\rm H}}\left(\frac{\Pp}{E_0}+\frac{1-\Pp}{E_\alpha}\right)\dep\\ 
\label{eq:evolb} \frac{dT_{\rm gas}}{dz}=&\frac{1}{1+z}\Big[2\Tg - \gamma_{_{\rm C}}\left(T_{\rm CMB}(z)-\Tg\right)\Big]+\\
\nonumber &-\frac{1}{(1+z)H(z)}\frac{1+2x_e}{3n_{\rm H}}\frac{2}{3\left(1+x_e+f_{\rm He}\right)}\dep \,.
\end{align}
\end{subequations}
The standard cosmological evolution is encoded in the first line of each equation where $\alpha_{\rm H}$ is the case-B recombination coefficient \cite{1989agna.book}, $\beta_{\rm H}$ the photoionization coefficient, $C\equiv\langle n_e^2\rangle/\langle n_e\rangle^2$ the clumping factor, $E_0=13.6 \eV$ the ground state binding energy of the hydrogen and $E_\alpha=3/4 E_0$ the energy associated with the Lyman-$\alpha$ transition. The Peebles $\Pp$ coefficient gives the probability for an hydrogen atom in the $n=2$ state to decay in the ground state before being ionized \cite{Peebles:1968ja}:
\be 
\Pp=\frac{1+K_{\rm H}\Lambda_{\rm H}n_{\rm H}(1-x_e)}{1+K_{\rm H}\left(\Lambda_{\rm H}+\beta_{\rm H}\right)n_{\rm H}(1-x_e)}\,,
\ee
where $\Lambda_{\rm H}\simeq 5.5 \times 10^{-15} \eV^2$ is the decay rate for the $2s$ state and $K_H=\pi^2/(E_\alpha^3 H(z))$. The effect of Compton scattering between CMB photons and free electrons is encoded in the dimensionless coefficient $\gamma_{_C}$:
\be
\gamma_{_C}=\frac{8\sigma_{\rm T}a_RT_{\rm CMB}^4}{3 H m_e }\frac{x_e}{1+x_e+f_{\rm He}}\,,
\ee
where $\sigma_T\simeq1.7\times10^3 \GeV^{-2}$ is the cross section for Thomson scattering, $a_R\simeq0.66$ is the radiation constant, $m_e$ the electron mass and $f_{\rm He}\equiv n_{\rm He}/n_b\simeq0.22$ the number fraction of Helium.

The second line in equations \eqref{eq:evola}-\eqref{eq:evolb} encodes the effect of energy injection due to DM decays. We followed the \textquotedblleft SSCK" prescription, in which a fraction $(1-x_e)/3$ of the deposited energy goes into ionization while a fraction $(1+2x_e)/3n_{\rm H}$ heats the plasma \cite{Chen:2003gz,SSCK}. In the analysis we used the energy deposition functions computed in \cite{1211.0283} and assumed a unit value for both $f_{\rm DM}$ and the clumping factor $C$ \cite{Chen:2003gz,MiraldaEscude:1998qs}. The mass dependence of the bounds only enters through the energy deposition functions $f(z,M_{\rm DM})$. 

Solving eq.\eqref{eq:evol} and assuming $T_{\rm S}=T_{\rm gas}$ we derive the maximum amplitude for the signal at $z=17.2$. Requiring that the 21 cm spectrum presents an absorption feature with $\delta T_b<-100\mK$ or $\delta T_b<-50\mK$, as suggested by the EDGES observation, we get the bounds reported in Fig. \ref{fig:bounds} and Fig. \ref{fig:all}. For DM in the mass range $1\MeV\div10\TeV$ decaying into $e^+e^-$, our bounds constrain $\tau_{_{\rm DM}}>10^{25}\s$. This constraint is significantly stronger than the one derived from observations of the CMB power spectrum \cite{1502.01589,1610.06933} and competitive with those coming from indirect searches \cite{1309.4091,1510.00389}. For the decay channel into photons the bounds become less stringent but still stronger than the ones from CMB. 
In Fig. \ref{fig:all} we report a compilation of bounds in the mass range $1\GeV\div 10\TeV$ for different decay channels. For each channel, we use the secondary electron and photon energy spectra provided in \cite{1012.4515,1009.0224}. 

\section{Comparison with the case of DM annihilations}
\label{sec:comparison}

In the case of DM annihilation the energy deposition rate is given by:
\be
\left(\frac{dE}{dVdt}\right)_{\rm deposited}=B(z) \, f^2_{_{\rm DM}}f\left(z,M_{\rm DM}\right) \rho^2_{_{\rm DM,0}}\, \frac{\langle \sigma v\rangle}{M_{\rm{DM}}} (1+z)^6 \,.
\ee
We note that for DM annihilations this rate depends on the small-scale inhomogeneities of the matter distribution through the cosmological boost factor $B(z)$. This introduces astrophysical uncertainties (related to our poor knowledge of the halo mass function, DM density profile, etc.) in the bounds on DM annihilations derived using the 21 cm line observation. Conversely, the energy injection rate \eqref{eq:energy_inj_decay} does not depend on the small-scale matter distribution; therefore bounds on the DM lifetime do not contain such astrophysical uncertainties.  

Moreover, in the case of DM decays the energy injection scales as $(1+z)^3$ while in the case of DM annihilation it scales as $(1+z)^6$. Therefore, bounds derived observing the 21 cm line (sensitive to processes happening around $z\sim20$) can improve the ones derived from CMB observables (sensible to processes happening around $z\sim 1000$) much more for DM decays than for annihilations.

Bounds on DM annihilation coming from the 21 cm line observation  were given in \cite{1803.03629}. In that work the authors use the effective $z$-independent absorption functions, $f_{\rm eff}$, provided by \cite{1506.03811,1506.03812} (i.e. they assume on-the-spot absorption)\footnote{An effective absorption coefficient could be defined also for the study of the dark ages. However, the use of the ones provided in \cite{1506.03811,1506.03812} is not justified here since they have been computed for processes taking places at much higher redshifts.}.  As shown in \cite{1603.06795}, this approximation breaks down at low redshifts where it is crucial to take into account the redshift dependence of both $f(z)$ and the boost factor $B(z)$\footnote{Notice that the deposition efficiency $f(z)$ for annihilations is different from the one for decays \cite{1604.02457}.}. The on-the-spot approximation overestimates the energy deposition for $z\gtrsim40$ while it underestimates it for lower redshifts. Thanks to this partial compensation, the results of \cite{1803.03629} turn out to be of the same order of the ones obtained using the correct $f(z)$ in the low mass regime ($M_{\rm DM}\lesssim 10\GeV$)\footnote{In the revised version of \cite{1803.03629} delayed energy deposition has been properly taken into account.}.

\section{Conclusions}
By neglecting astrophysical sources of IGM heating and considering the effect of DM decay, we derived conservative bounds on the lifetime of DM which are independent of astrophysical parameters.
The results, summarised in Fig.~\ref{fig:bounds} and Fig. \ref{fig:all}, are stronger than the limits derived by CMB observations and competitive with indirect probes of DM decays.

These bounds rest on the assumption that there are no non-standard sources of gas cooling (\emph{e.g.} DM-baryon scattering) nor additional sources of 21 cm photons that modify the tail of the CMB spectrum.
Differently from the case of DM annihilations, the results are independent of the DM density inhomogeneities and thus they are not affected by the uncertainties on the DM distribution on small scales.

\vspace{1.5cm}

\noindent While this work was been completed, references \cite{1803.09390,1803.09739} appeared on the arXiv. They derive similar bounds on the DM lifetime which are consistent with our results. This work adopted a slightly different point of view regarding the treatment of astrophysical parameters and the inclusion of non-standard sources of gas cooling or excess background radiation. In particular \cite{1803.09739} discusses how the bounds are modified in the presence of extra cooling.

\subsection*{Aknowledgments}
We thank R. Contino and A. Mesinger for useful discussions.

\end{document}